\documentclass{elsart3p}
\usepackage{epsfig}
\usepackage{color}

\usepackage{amsmath} 
\usepackage{graphicx} 
\usepackage{verbatim} 
\usepackage{color} 
\usepackage{subfigure} 
\usepackage{longtable}
\usepackage{dcolumn}
\usepackage{epsfig}
\newcommand{\ee}{$e^{+}e^{-}$}

\newcommand{\gevu}{GeV/$u$  }

\newcommand{\gevcc}{GeV/$c^{2}$}

\newcommand{\gevc}{GeV/$c$}

\newcommand{\gev}{GeV}
\newcommand{\mev}{MeV}

\begin{document}
\begin{frontmatter}

\title{Searching a Dark Photon with HADES}

\collaboration{HADES Collaboration}

\author[7]{G.~Agakishiev},
\author[3]{A.~Balanda},
\author[18]{D.~Belver},
\author[7]{A.~Belyaev},
\author[9]{J.C.~Berger-Chen},
\author[2]{A.~Blanco},
\author[10]{M.~B\"{o}hmer},
\author[16]{J.~L.~Boyard},
\author[18]{P.~Cabanelas},
\author[7]{S.~Chernenko},
% ...

\vspace*{0.3cm}

\address[1]{Istituto Nazionale di Fisica Nucleare - Laboratori Nazionali del Sud, 95125~Catania, Italy}
\address[2]{LIP-Laborat\'{o}rio de Instrumenta\c{c}\~{a}o e F\'{\i}sica Experimental de Part\'{\i}culas , 3004-516~Coimbra, Portugal}
\address[3]{Smoluchowski Institute of Physics, Jagiellonian University of Cracow, 30-059~Krak\'{o}w, Poland}
\address[4]{GSI Helmholtzzentrum f\"{u}r Schwerionenforschung GmbH, 64291~Darmstadt, Germany}
\address[5]{Technische Universit\"{a}t Darmstadt, 64289~Darmstadt, Germany}
\address[6]{Institut f\"{u}r Strahlenphysik, Helmholtz-Zentrum Dresden-Rossendorf, 01314~Dresden, Germany}
\address[7]{Joint Institute of Nuclear Research, 141980~Dubna, Russia}
\address[8]{Institut f\"{u}r Kernphysik, Goethe-Universit\"{a}t, 60438 ~Frankfurt, Germany}
\address[9]{Excellence Cluster 'Origin and Structure of the Universe' , 85748~Garching, Germany}
\address[10]{Physik Department E12, Technische Universit\"{a}t M\"{u}nchen, 85748~Garching, Germany}
\address[11]{II.Physikalisches Institut, Justus Liebig Universit\"{a}t Giessen, 35392~Giessen, Germany}
\address[12]{Istituto Nazionale di Fisica Nucleare, Sezione di Milano, 20133~Milano, Italy}
\address[13]{Institute for Nuclear Research, Russian Academy of Science, 117312~Moscow, Russia}
\address[14]{Institute of Theoretical and Experimental Physics, 117218~Moscow, Russia}
\address[15]{Department of Physics, University of Cyprus, 1678~Nicosia, Cyprus}
\address[16]{Institut de Physique Nucl\'{e}aire (UMR 8608), CNRS/IN2P3 - Universit\'{e} Paris Sud, F-91406~Orsay Cedex, France}
\address[17]{Nuclear Physics Institute, Academy of Sciences of Czech Republic, 25068~Rez, Czech Republic}
\address[18]{LabCAF. F. F\'{\i}sica, Univ. de Santiago de Compostela, 15706~Santiago de Compostela, Spain} 
\vspace*{0.3cm}

\address[a]{Also at Lawrence Berkeley National Laboratory, ~Berkeley, USA}
\address[b]{Also at ISEC Coimbra, ~Coimbra, Portugal}
\address[c]{Also at ExtreMe Matter Institute EMMI, 64291~Darmstadt, Germany}
\address[d]{Also at Technische Universit\"{a}t Dresden, 01062~Dresden, Germany}
\address[e]{Also at Dipartimento di Fisica, Universit\`{a} di Milano, 20133~Milano, Italy}
\address[f]{Also at Frederick University, 1036~Nicosia, Cyprus}
\address[g]{Also at Dipartimento di Fisica Generale and INFN, Universit\`{a} di Torino, 10125~Torino, Italy}
\vspace*{0.3cm}

\corauth[cor]{Corresponding authors: M.Gumberidze@gsi.de, R.Holzmann@gsi.de}

\date{\today}

\begin{abstract}
We present a search for the $e^+e^-$ decay of a hypothetical dark photon, also names $U$ vector boson, in 
inclusive dielectron spectra measured by HADES in the p (3.5~GeV) + p, Nb reactions, as well
as the Ar (1.756 \gevu  ) + KCl reaction.  An upper limit on the kinetic mixing parameter squared ${\epsilon^{2}}$
at $90\%$~CL has been obtained for the mass range $M_{U} = 0.02 - 0.55$~\gevcc and is compared with
the present world data set.  For masses 0.03 - 0.1~\gevcc, the limit has been lowered with respect
to previous results, allowing now to exclude a large part of the parameter region favoured by
the muon $g-2$ anomaly.  Furthermore, an improved upper limit on the branching ratio of $2.3 \times 10^{-6}$ has been
set on the helicity-suppressed direct decay of the eta meson, $\eta \rightarrow e^+e^-$, at 90\%~CL.

\end{abstract}

\begin{keyword}
dark photon, hidden sector, dark matter, rare eta decays
\PACS{13.20.-v, 25.40.-h, 95.35.+d}
\end{keyword}
\end{frontmatter}

\section{Introduction}
Observations of the cosmic electron and/or positron flux by ATIC \cite{atic}, PAMELA \cite{pamela},
HESS \cite{hess1,hess2}, Fermi \cite{fermi}, and recently the AMS02 collaboration \cite{ams02} have
revealed an unexpected excess at momenta above 10~\gev, in particular in the positron fraction $e^+/(e^- + e^+)$.
These observations can not easily be reconciled in a consistent way with known astrophysical sources \cite{wimpdecays1}
and alternative theoretical explanations have therefore been put forward.  In particular, scenarios
in which the excess radiation stems from the annihilation of weakly interacting dark matter particles \cite{wimpdecays1,wimpdecays2}
might offer an enticing solution to this puzzle.  There is indeed compelling evidence from various astronomical
and cosmological observations \cite{darkmatter,pdg2012} that non-baryonic matter of some sort is responsible for
20-25\% of the total energy density in the Universe.  This so-called dark matter (DM) is assumed to be a relic
from the Big Bang making itself noticeable by its gravitational action on the large-scale cosmic structures.
To accomodate DM in elementary particle theory and to allow it to interact with visible matter,
it has been proposed to supplement the Standard Model (SM) with an additional sector characterized by another
$U(1)'$ gauge symmetry \cite{fayet80,boehm,fayet04,pospelov08}.  The corresponding vector gauge boson
--- called $U$~boson, $A'$, $\gamma'$, or simply dark photon --- would thereby mediate the annihilation
of DM particles into charged lepton pairs.  Indeed, from theoretical arguments a kinetic mixing of the $U(1)'$
and $U(1)$ symmetry groups would follow \cite{okun,holdom}, providing a natural connection between the dark
and SM sectors.  For that purpose, a mixing parameter $\epsilon$ has been introduced \cite{fayet80} relating
the respective coupling strengths ($\alpha`$) of the dark and SM photons to visible matter via $\epsilon^2 = \alpha'/\alpha$;
it is expected to be of order ${10^{-2}-10^{-8}}$ \cite{arkani}.  Also, the mass of the $U$~boson is thought to
remain well below 1~\gevc\ \cite{arkani}, resulting most likely in a small width $\Gamma_U$ $\ll$ 1~\mev\ \cite{bjorken,batell1,reece}.
This is of particular interest for experimental searches because a dark photon would appear in the data as
a rather narrow resonance. 

Through the $U(1) - U(1)'$ mixing term the $U$~boson would be involved in all processes which include real
or virtual photons \cite{reece}.  On the other hand, any search for a $U$ boson will have to deal with the large
irreducible background from standard QED radiative processes \cite{landsberg}.  In recent years, a number of
such searches have been conducted in various experiments done in the few-GeV beam energy regime, looking either
at $e^+e^-$ pair distributions produced in electron scattering \cite{mamiA1,apex} or in the electromagnetic decays
of the neutral pion \cite{sindrum,wasa} and the $\phi$ meson \cite{kloe-2a,kloe-2b}.  In particular, the latter
experiment exploited the hypothetical $\phi \rightarrow \eta + U\rightarrow 3\pi e^{-}e^{+}$ decay with the $\phi$ produced
in $e^+e^-$ collisions.  Reconstructing the $e^+e^-$ invariant-mass distribution tagged by fully identified $\eta$ mesons
in either of their two 3-pion decay channels, $\pi^0\pi^0\pi^0$ or $\pi^+\pi^-\pi^0$, a search for a narrow
$U \rightarrow e^+e^-$ signal was possible.  In a similar fashion the WASA-at-COSY experiment \cite{wasa}
has covered the mass range $M_U = 0.02-0.1$~\gevcc\ --by investigating decays of $\pi^0$ produced in proton-induced
reactions at 0.55~\gev\ beam energy.  By analyzing data obtained from high-flux neutrino production experiments at CERN,
regions in parameter space $\epsilon^2$ vs. $M_U$ corresponding to a long-lived $U$ have been excluded as well \cite{gninenko}.
Note finally, that from the very precisely measured value of the anomalous gyromagnetic factors $(g-2)$ of the
muon and electron \cite{muong2}, additional constraints are put on the allowed range of the mixing parameter
$\epsilon$ and the mass $M_U$ \cite{pospelov09,endo}.

Here, we present results of a search for a $U\rightarrow e^{-}e^{+}$ decay signal in inclusive dielectron
spectra obtained from 3.5~GeV proton-induced reactions on either a liquid hydrogen target or a solid niobium target,
as well as Ar (1.756~\gevu\ ) + KCl reaction.  The reconstructed dielectron invariant-mass distribution from those
reactions, as well as data on the respective inclusive $\pi^0$ and $\eta$ production have been published
elsewhere \cite{hades_p35p,hades_pNb,hades_pi0_eta,hades_arkcl}.  This paper is organized as follows:
In Section~2 we discuss the $e^+e^-$ decay signature of a hypothetical dark photon.  Section~3 presents
the HADES experiment and data analysis, Sec.~4 describes in detail our $U$-boson search, in Sec.~5 we give
a new upper limit on the direct $\eta$ decay, and, finally, in Sec.~6 we summarize our findings.


\section{The $\bf{U} \rightarrow \bf{e^+e^-}$ signature}

Unlike the experiments described in \cite{kloe-2a,kloe-2b,wasa}, HADES has measured inclusive
instead of exclusive dielectron production.  This means that the reconstructed $e^+e^-$ invariant-mass distribution $dN/dM_{ee}$
consists of a cocktail of contributions from different sources, mainly the electromagnetic decays of
$\pi^0$, $\eta$, and $\Delta$  \cite{hades_p35p}, and our search for signatures of a hypothetical
$U$ boson has to take this into account \cite{batell2}.

Let us estimate the $U$-boson yield by $N_{U} = \sum_{i} N^{(i)}_{U} $, where $N^{(i)}_{U}$ refers to
separable sources, such as $i=\pi^{0}, \eta $ and $\Delta$, with the virtual photon (i.e. dilepton)
replaced by a $U$.  We obtain the ratios of widths from data via 

\begin{subequations}

\begin{align}
\frac{\Gamma_{i \rightarrow \gamma U}}{\Gamma_{i \rightarrow \gamma\gamma}} = \frac{N^{(i)}_{U}}{N_{i}\,BR_{i \rightarrow \gamma \gamma}}, \\
\frac{\Gamma_{\Delta \rightarrow N U}}{\Gamma_{\Delta \rightarrow N \gamma}} = \frac{N^{\Delta}_{U}}{N_{\Delta}\,BR_{\Delta \rightarrow N \gamma}}, 
\end{align}
\end{subequations}

\noindent
where $i=\pi^{0}$ and $\eta$.  To get access to $\epsilon^2$, we use the expression

\begin{equation}
\frac{\Gamma_{i \rightarrow \gamma U}}{\Gamma_{i \rightarrow\gamma\gamma}} = 2\epsilon^{2}\left| F_{i}(q^{2}=M_{U}^{2}) \right|\frac{\lambda^{3/2}(m_{i}^{2},m_{\gamma}^{2},M_{U}^{2})}{\lambda^{3/2}(m_{i}^{2},m_{\gamma}^{2},m_{\gamma}^{2})}.
\end{equation}

\noindent
Here, $\lambda$ is the standard triangle function for relativistic kinematics and $F_{i}(q^2)$ is the electromagnetic transition form factor.
Furthermore, for on-shell photons ($m_{\gamma}^{2} = 0$), one gets

\begin{equation}
\frac{\lambda^{3/2}(m_{\eta}^{2},0,M_{U}^{2})}{\lambda^{3/2}(m_{\eta}^{2},0,0)}=(1-\frac{M_{U}^{2}}{m_{\eta}^{2}})^{3}.
\label{massFac_eta}
\end{equation}

\noindent

Note that, as the $\Delta$ is a broad state, the decay width $\Gamma_{\Delta \rightarrow N U}$ has to be
averaged over the $\Delta$ mass distribution $A(m_{\Delta})$, assumed to be described by a Breit-Wigner shape
of width $\Gamma=117$~\mev\ (see \cite{pluto} for details):

\begin{multline}
\frac{\Gamma_{\Delta\rightarrow N U}}{\Gamma_{\Delta\rightarrow N\gamma}} = \\
\epsilon^2 \int A(m_{\Delta}) \left| F_{\Delta}(M_{U}^2) \right|\frac{\lambda^{3/2}(m_{\Delta}^2,m_N^2,M_{U}^2)}{\lambda^{3/2}(m_{\Delta}^2,m_N^2,0)} \mathrm{d}m_{\Delta}.
\label{gammaDelta_b}
\end{multline}

One has to consider furthermore that, as the $\eta$ and $\Delta$ decays give access to masses larger than
the $\mu^+\mu^-$ threshold at $2 m_{\mu} = 0.21$~\gevcc\ , the observed $U$ signal has to be corrected for the
branching fraction into $e^+e^-$, that is $BR_{ee} = BR_{U\rightarrow e^+e^-}$ \cite{batell1}:

\begin{equation}
BR_{ee} = \Gamma_{ee}/\Gamma_{tot} = \frac{\Gamma_{ee}}{\Gamma_{ee} + \Gamma_{\mu\mu} + \Gamma_{had}}.
\label{BRee_gamma}
\end{equation}

\begin{figure}[ht!]
  \begin{center}
  \mbox{\epsfig{width=0.70\linewidth, height=0.70\linewidth, figure=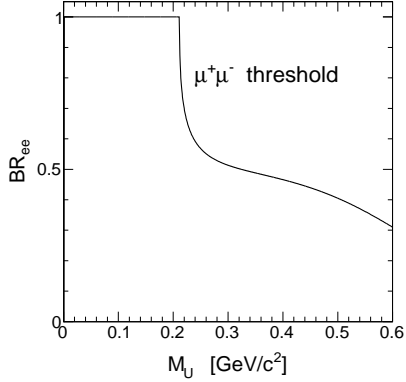}}
  \vspace*{-0.2cm}
  \caption[] {Assumed branching ratio $BR_{ee}$ of a hypothetical $U$ boson into an
    $e^+e^-$ pair as a function of mass $M_U$ according to Eq.~(\ref{BRee_mass}).
    }
  \end{center}
  \vspace*{-0.2cm}
  \label{BRee}
\end{figure}

Assuming lepton universality, that is $\Gamma_{\mu\mu} = \Gamma_{ee}$ for $M_U \gg 2 m_{\mu}$, and estimating
the hadronic decay width by
$R(\sqrt{s}) = \sigma_{e^+e^-\rightarrow hadrons}/\sigma_{e^+e^-\rightarrow \mu^+\mu^-}$ factor
(taken from \cite{pdg2012}), such that $\Gamma_{had} = R(M_U)\; \Gamma_{\mu\mu}$,
the branching relevant for our search is given by



\noindent

\begin{equation}
BR_{ee} = \frac{1}{1 + \sqrt{1 - \frac{4 m_{\mu}^2}{M_U^2}} \; (1 + \frac{2 m_{\mu}^2}{M_U^2}) \; [1 + R(M_U)]}.
\label{BRee_mass}
\end{equation}

Figure \ref{BRee} exhibits $BR_{ee}$ as a function of $M_{U}$. 

\noindent
Plugging all together we obtain

\begin{equation}
\begin{split}
N_{U \rightarrow ee} &= N_{U\rightarrow ee}^{\eta} + N_{U\rightarrow ee}^{\pi^0} + N_{U\rightarrow ee}^{\Delta} \\ 
&= \epsilon^2 \, BR_{ee} \, [ 2 N_{\eta} \, BR_{\eta\rightarrow\gamma\gamma} \,|F_{\eta}|^2 \,(1-M_U^2/m_{\eta}^2)^3 \\
&+ 2 N_{\pi^0} \, BR_{\pi^0\rightarrow\gamma\gamma}\, |F_{\pi^0}|^2 \,(1-M_U^2/m_{\pi^0}^2)^3 \\
&+ N_{\Delta} \, BR_{\Delta\rightarrow N \gamma}  \times\\
&~~\int A(m_{\Delta}) \,|F_{\Delta}|^2 \frac{\lambda^{3/2}(m_{\Delta}^2,m_N^2,M_U^2)}{\lambda^{3/2}(m_{\Delta}^2,m_N^2,0)}
 \,\mathrm{d}m_{\Delta} ] \\ 
&= \epsilon^2 \, BR_{ee} \, L(M_U),
\end{split}
\label{nU}
\end{equation}

\noindent
where $L(M_U)$ assembles all kinematic factors and source parameters in Eq.~(\ref{nU}). 

If no actual $U$ signal is observed and only an upper limit on the $U$ multiplicity can be given,
it yields accordingly an upper bound on $\epsilon^{2}$ as a function of $M_{U}$.

Note that our approach is based on the following assumptions: (i) $i=\pi^{0}$, $\eta$, and $\Delta$ saturate the sum over all $U$-boson sources,
(ii) the estimate of $BR_{U \rightarrow e^{+}e^{-}}$ is sufficiently accurate, (iii) the parametrization of the transition form factors
$\left|F_{\pi^0}(q^{2})\right|=1+0.032\; q^2/m_{\pi^0}^2$ \cite{pdg2012}
and $\left|F_{\eta}(q^{2})\right|=(1-\frac{q^{2}}{\Lambda^{2}})^{-1}$ with ${\Lambda=0.72}$~GeV \cite{berghauser,NA60} are
accurate enough, (iv) the spectral distribution of the $\Delta$ in Eq.~(\ref{gammaDelta_b}) is correct,
(v) the use of $\left|F_{\Delta}(q^{2})\right|=1$ does not alter the result, since an experimental form factor is
not known (although \cite{pena} argues on a weak $q^{2}$ dependance),
(vi) uncertainties in the estimates of the $\Delta$ multiplicities by $N_{\Delta} = 3/2 N_{\pi^0}$ are of
minor importance due to the small value of $BR_{\Delta \rightarrow N \gamma}=0.006$ compared with
$BR_{\eta\rightarrow\gamma\gamma}=0.393$, $BR_{\pi^0\rightarrow\gamma\gamma}=0.988$ \cite{pdg2012}. 

\section{The HADES experiment}

The high-acceptance dielectron spectrometer HADES operates at the
GSI Helmholtzzentrum f\"{u}r Schwerionenforschung in Darmstadt, where it
uses the beams from the heavy-ion synchrotron SIS18 in the few-GeV beam-energy
range.
A detailed description of the set-up can be found in~\cite{hades_tech}.

In the experiments discussed here a proton beam with a kinetic energy of
E$_p$ = 3.5~\gev\ and an average intensity of about  $2\times10^{6}$ particles per second
was used to bombard either a solid 12-fold segmented niobium target (with 2.8\% nuclear
interaction probability) \cite{hades_pNb} or a liquid hydrogen target (1\% interaction probability) \cite{hades_p35p}.
In both experiments events were registered if at least three charged-particle hits
were registered in the HADES time-of-flight wall (LVL1 trigger) and those events were actually
recorded in case at least one electron or positron candidate was detected (LVL2 trigger).
In the third experiment, a 4-fold segmented potassium chloride (KCl) target was bombared
with a $^{40}$Ar beam (kinetic beam energy of 1.75 \gevu), the LVL1 trigger requiring at least 16 hits in the TOF wall \cite{hades_arkcl}.

\begin{figure}[!htb]

\mbox{\epsfig{figure={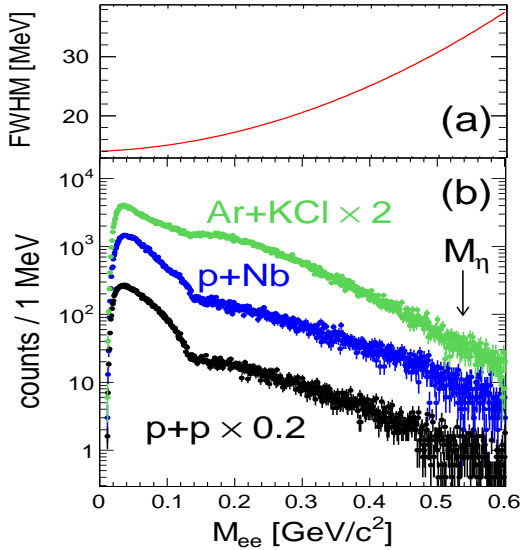}, width=1.\linewidth,height=1.0\linewidth}}
\vspace*{-0.3cm}

  \caption[]{(Color online)
  (a) Dielectron mass resolution (FWHM) as a function of the \ee\ invariant mass obtained from a Monte-Carlo
  simulation.  
  (b) Measured inclusive \ee\ invariant-mass distributions for 3.5 GeV p+p and p+Nb reactions, respectively and 1.756 \gevu Ar+KCl
  reactions in the HADES geometrical acceptance with single lepton momenta $p_e>0.05$~\gev\ and pair opening
  angles $\theta_{e+e-}>9^\circ$.  Error bars are statistical only.
  The arrow indicates the position where a direct $\eta$ decay peak would appear ($M_{\eta} = 0.548$~\gevcc\ ).
  Note that the peak is not visible, therefore an upper limit can be extracted at the expected position, see Section V.
}
  
  \label{signal_res}

\end{figure}

In the data analysis, electrons and positrons were identified by applying
selection cuts to the RICH, pre-shower and energy-loss signals.  The particle momenta
were obtained by tracking the charged particles through the HADES magnetic field; the latter were
combined two-by-two to fully reconstruct the 4-momentum of \ee\ pairs.  A detailed description of
this analysis is given in \cite{hades_arkcl,hades_tech}.  Figure~\ref{signal_res} shows the
resulting reconstructed invariant-mass distributions from the three reactions.  As all reactions were
investigated with \texttt{¥}he same setup, the detector acceptances and efficiencies were comparable.
Still, as discussed in the next section, we have conducted separate searches in the three data sets
and join the results in the end.

The production cross-sections (or multiplicities) of $\eta$ and $\pi^0$ mesons have been published 
in \cite{hades_p35p,hades_pi0_eta,hades_arkcl} for the p+p, p+Nb, and Ar+KCl experiments, respectively.
Recalculated total numbers of mesons ($N_{\eta}$ and $N_{\pi^0}$) produced in those experiments are listed
in Tab.~\ref{pp_pNb_mesons}.  For the $\Delta$ resonance the factor 3/2 in $N_{\Delta} = 3/2 N_{\pi^0}$
has to be seen as an extreme, assuming that all pion production is mediated
by $\Delta$ decays, whereas model calculations typically favor smaller numbers \cite{GiBUU}.  In fact,
because of the small electromagnetic branching $BR_{N\gamma}$ of the $\Delta$ resonance, its contribution to
dark photon production is small compared to the $\pi^0$ and $\eta$.

\begin{table}[htb]
  \begin{center}
  \begin{tabular}{ l| c| c | c}
Reaction  & ~~$N_{LVL1}$~~ & ~~~~$N_{\pi^0}$~~~ & ~~~~$N_{\eta}$~~~~\\
  \hline \hline
 ~p+p     &  $3.0 \times 10^9$ & $2.5 \times 10^9$ & $1.5 \times 10^8$ \\
 ~p+Nb    &  $7.7 \times 10^9$ & $5.9 \times 10^9$ & $3.0 \times 10^8$ \\
 ~Ar+KCl~ &  $2.2 \times 10^9$ & $7.7 \times 10^9$ & $1.9 \times 10^8$ \\
  \end{tabular}
\caption[]{Total number of triggered events $N_{LVL1}$  as well as number of $\pi^0$ ($N_{\pi^0}$) and $\eta$ ($N_{\eta}$) mesons
   produced in the HADES p+p, p+Nb, and Ar+KCl experiments, respectively.  The latter has been recalculated from
   the production data published in \cite{hades_p35p,hades_pi0_eta,hades_arkcl}.  
   Experimental uncertainties on the meson yields are of order 15 -- 25\%.}
  \end{center}
    \label{pp_pNb_mesons}
\end{table}

\begin{figure*}

\mbox{\epsfig{figure={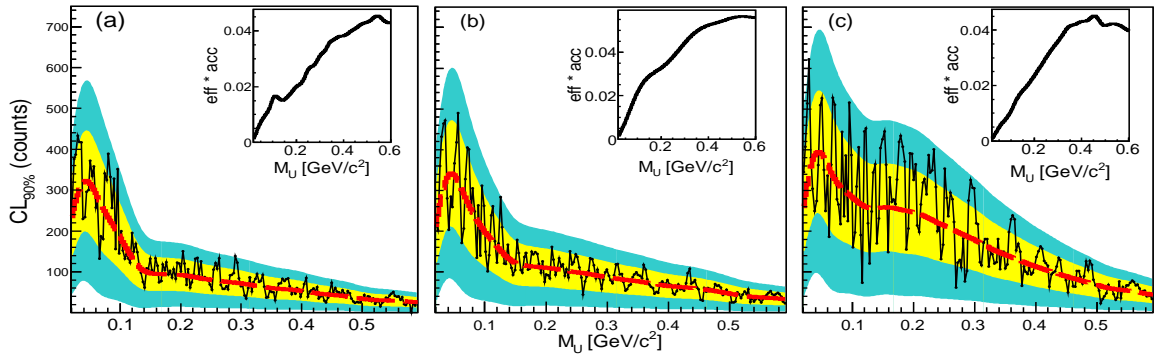}, width=0.95\linewidth,height=0.31\linewidth}}
\vspace*{-0.3cm}
  \caption[]{(Color online)
  Extracted 90\% CL upper limits on a narrow $U\rightarrow e^+e^-$ signal as function of $M_{ee}$ for p+p (a),
  p+Nb (b), and Ar+KCl (c) uncorrected HADES data (symbols).  The computed experimental sensitivity (median UL) is shown
  as pink dashed curve and its error bands are given in yellow ($\pm1\sigma$) and cyan ($\pm2\sigma$).
  The inserts show the respective product of pair efficiency and pair acceptance, $eff \times acc$ vs. $M_{ee}$.
  }

  \label{UL}

\end{figure*}

\section{The $\bf{U}$-boson search}

As discussed above, the search for the $U$ boson can be performed with HADES using all electromagnetic
decays typically populated in few-GeV hadronic interactions, that is mostly $\pi^0 \rightarrow \gamma U$,
$\eta \rightarrow \gamma U$, and $\Delta \rightarrow N U$, followed by $U \rightarrow e^+e^-$.
In contrast to previous experiments \cite{wasa,kloe-2a,kloe-2b} focussing on a specific decay channel,
our search is based on the inclusive measurement of all \ee\ pairs produced in a given mass range. 
An irreducible background due to the respective Dalitz decays of the $\pi^0$, $\eta$, and $\Delta$
is always present.  Indeed, because of their very similar decay kinematics, the latter sources
cannot be discriminated from a $U$-boson signal via analysis cuts.  Therefore, we have to search for
a peak structure on top of a smoothly varying continuum.
 Because of the expected long lifetime of the new particle the width of such a peak will be determined by the detector resolution.
The upper frame of Fig.~\ref{signal_res} (a) shows the mass resolution obtained from a GEANT3-based
Monte Carlo of \ee\ decays detected in the HADES detector.  The calculated peak width increases smoothly
with pair mass from about 15~\mev\ (fwhm) in the $\pi^0$ region to about 30~\mev\ at the
$\eta$ mass of 0.55~\gevcc\ . 

The present analysis is based on the raw dilepton mass spectra, exhibited on Fig.~\ref{signal_res} (b) i.e.~spectra not corrected for efficiency
and acceptance. The low invariant-mass region of the spectra (M$_{ee}<0.13$~\gevcc\ ) is dominated by $\pi^0$ Dalitz decays,
at intermediate masses (0.13~\gevcc\  $< M_{ee} < $  0.55~\gevcc\ ), $\eta$ and $\Delta$ Dalitz decays prevail, and
the high-mass region is populated mostly by low-energy tails of vector-meson decays \cite{hades_p35p,hades_pNb}.
However, as the electromagnetic decay branching ratios decrease with increasing particle mass,
resulting in low sensitivity, we restrict our search to $M_U<0.6$~\gev\ .

The sensitivity of the experiment for observing a peak-like $U \rightarrow e^+e^-$ mass signal
depends evidently on various factors: the geometric acceptance of HADES for these decays, on the combined
detection and reconstruction efficiency of the \ee\ signal, on its mass resolution, and on the
signal-over-background ratio $S/B$.  The latter one is not only given by the purity of the pair
signal $per~se$, it also reflects the amount of uncorrelated lepton pairs constituting the so-called
combinatorial background (CB).  Whereas a high purity of the dielectron signal is guaranteed by the overall
good quality of the HADES lepton identification, the CB can not be fully suppressed by analysis cuts.
Although its contribution can be determined quite accurately either by event-mixing techniques or
from the yields of same-event like-sign pairs \cite{hades_tech}, it is always part of the total reconstructed
pair yield and hence does contribute to the Poisson fluctuations of the latter. 






Our search for a narrow resonant state in the \ee\ mass distributions has been conducted in the
following way. The $dN/dM_{ee}$ spectrum (Fig.~\ref{signal_res} (b)), measured in either of the analyzed reactions,
was fitted piece-wise with a model function consisting of a $5^{th}$-order polynomial and a Gauss peak
of fixed position $M_{ee}$ and fixed width $\sigma(M) = fwhm/2.35$ (from the simulation shown
in Fig.~\ref{signal_res}(a)). The adjustment was done by sliding a fit window of width $\pm 4\sigma(M)$
over the spectrum in steps of 3~\mev.  In each such step, the fit delivered a parameterization
of the local background in presence of a possible gaussian signal of given width $\sigma(M)$.
This analysis shows that no significant
peak is present in our data (see also Fig.~\ref{signal_res} (b)).  Consequently, a statistical likelihood-based test must be performed
to determine at a given Confidence Level (CL) an upper limit (UL) for a possible $U$-boson signal \cite{cowan}.
Such tests are usually based on the profile likelihood ratio computed as a function of the signal
strength $S$ in presence of so-called nuisance parameters, e.g. the known (or estimated)
background yield, the geometric acceptance, the detector and reconstruction efficiencies, and any
overall normalization factors.  As, in our case, background and \ee\ efficiency corrections are needed
to extract an absolute signal yield, and as both are known with limited accuracy only, we have used
the extended profile likelihood method proposed by Rolke, Lopez and Conrad \cite{rolke} to compute the UL at a confidence level CL~=~90\%.

In our search, we have hence integrated the total observed dilepton yield as well as the adjusted smooth
background over an interval $\pm 1.5\sigma(M)$ centered at each examined mass $M_U$.  Note that the chosen
integration window assures 90\% coverage of any hypothetical narrow signal at that mass.  As we deal with
sizable experimental yields, in the range of a few 100 to a few 1000 counts per inspected mass bin,
we have applied the Root implementation \cite{root} of the procedure \cite{rolke} assuming a gaussian error on
the background as well as on the product of the acceptance and efficiency corrections ($acc \times eff$).
The gaussian background error was provided by the polynomial least-square fit and the systematic
error on all correction factors was determined to be 15\%.  This value encompasses in particular the
error on the published particle production cross sections and electromagnetic branching ratios $BR_{\gamma\gamma}$.

The resulting upper limits, expressed as detectable counts, are shown in Fig.~\ref{UL} for the
mass range covered in this experiment, i.e. 0.02 -- 0.55~\gevcc\ .  This figure also shows the expected
sensitivity of our experiments, determined by running a Monte Carlo simulation in which the experimental
mass spectrum was resampled channel by channel many times.  In each such an iteration, the UL has been
re-evaluated with the ``zero-signal'' hypothesis, i.e. assuming $S=0$.  This way, after 10,000 iterations,
the median and standard deviation of the generated UL distributions could be computed as a function
of pair mass \cite{cowan}.  The experimental sensitivity can in fact be characterized as the median
significance with which a non-zero result of the search (at $S=0$) can be rejected at a given CL.
Figure~\ref{UL} shows the obtained median UL together with its respective $\pm1\sigma$ and $\pm2\sigma$ error bands.
Assuming a normally distributed UL, 68\% (95\%) of the sampled UL should be contained within the $\pm1\sigma$
($\pm2\sigma$) envelopes.  Note that the UL determined from the actual data sets do fluctuate about the
calculated median while staying indeed within the 
\vspace{0.5cm}
expected corridors with roughly the expected rate.

\begin{figure}[t!]

\mbox{\epsfig{figure={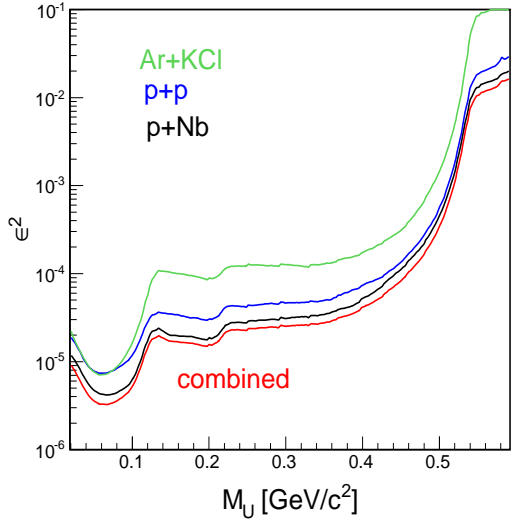}, width=1.\linewidth,height=1.0\linewidth}}
\vspace*{-0.3cm}

  \caption[]{(Color online)
   Exclusion plot at 90\% CL on $\epsilon^{2}$ as function of $M_U$
   from the analyses of HADES in the reactions p(3.5 \gev) + Nb , as well as Ar (1.756\gevu) + KCl.  Also shown is the
   combined UL computed with Eq.~(\ref{limit_add}).
}
  \label{UL_eps1}

\end{figure}

The inserts in Fig.~\ref{UL} show, as a function of mass, the pair efficiency and acceptance
correction factor, $eff \times acc$, obtained from detailed simulations.  After having
corrected the median UL for this factor, Eq.~(\ref{nU}) was used to compute a corresponding
upper limit $UL(\epsilon^2)$ on the relative coupling strength $\epsilon^2$ of a hypothetical dark vector boson.
Figure~\ref{UL_eps1} shows the $UL(\epsilon^2)$ as a function of $M_U$ obtained from the three data sets separately.
Evidently, the p+Nb data provide the strongest constraint.  However, as the three data sets are of
comparable statistical quality and result hence in upper limits of similar magnitude,
it is natural to join them into a combined upper limit \cite{helene}.  Since all experiments
having been executed under very similar conditions, we use the following statistics-driven ansatz:

\begin{equation}
 UL_{(1+2+3)} = \sqrt{(UL_{(1)}^{-2} + UL_{(2)}^{-2}  + UL_{(3)}^{-2} )^{-1}}.
 \label{limit_add}
\end{equation}

The combined upper limit $UL_{(1+2+3)}$ is overall about 10~to~20\% lower than the p+Nb
value taken alone.  This is indeed expected from the moderate increase in pair statistics
achieved by cumulating the data from all experiments and is consistent with
a $UL\propto 1/\sqrt{N}$ behavior.

\begin{figure}[ht!]

\mbox{\epsfig{figure={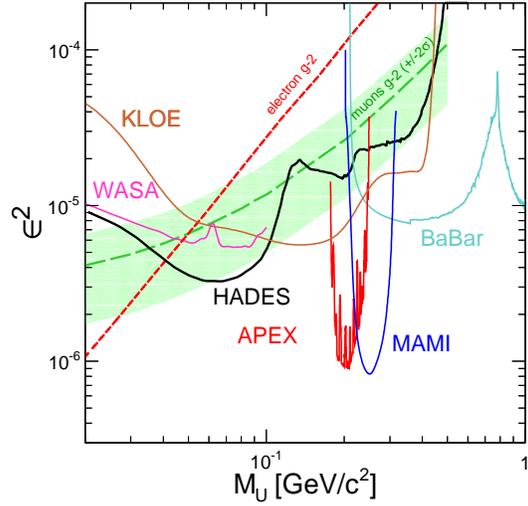}, width=1.\linewidth,height=1.0\linewidth}}
\vspace*{-0.3cm}

  \caption[]{(Color online)
   The 90\% CL upper limit on $\epsilon^{2}$ versus the $U$-boson mass obtained from the combined analysis
   of HADES data (solid black curve).  This result is compared with existing limits from the MAMI/A1,
   APEX, BaBar, WASA, and KLOE-2 experiments, as well as with the $g-2$ constraints (see the text for citations).}
  \label{UL_eps2}

\end{figure}

Finally, in Fig.~\ref{UL_eps2} we show the HADES result together with a compilation of limits from the
searches conducted by BaBar~\cite{BaBar,bjorken,reece}, KLOE-2~\cite{kloe-2a,kloe-2b}, APEX~\cite{apex},
WASA at COSY \cite{wasa}, and A1 at MAMI~\cite{mamiA1}.  At low masses ($M_U<0.1$~\gevcc) we clearly improve
on the recent result obtained by WASA \cite{wasa}, excluding now to a large degree the parameter range
allowed by the muon $g-2$ anomaly (preediction with 2$\sigma$ interval is shown on the Fig.~\ref{UL_eps2}).  At higher masses, the sensitivity of our search is compatible with,
albeit somewhat lower than the combined KLOE-2 analysis of $\phi$ decays.  Our data probe, however, the $U$-boson
coupling in $\eta$ decays and add hence complementary information.  At masses above the $\eta$ mass,
the inclusive dilepton spectrum is fed by $\Delta$ (and to some extent heavier baryon resonance) decays
which offer only small sensistivity, partly due to the small electromagnetic branching ratio
($BR_{N\gamma} \simeq 10^{-3} - 10^{-2}$) and partly due to the decreasing $BR_{U\rightarrow ee}$ at high $M_U$.

\section{UL on the rare decay $\bf{\eta \rightarrow e^+e^-}$}

\begin{figure}[ht]

\mbox{\epsfig{figure={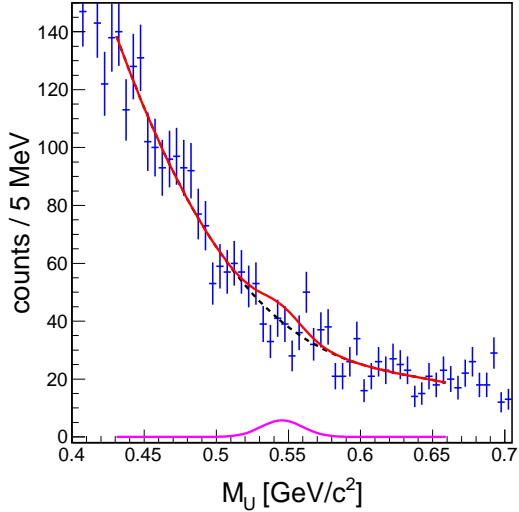}, width=1.\linewidth,height=1.0\linewidth}}
\vspace*{-0.3cm}

  \caption[]{(Color online)
Zoom into the $\eta$ peak region of the invariant-mass distribution of \ee\ pairs reconstructed
in the p(3.5GeV) + Nb reaction.  The data is fitted with a polynomial (dashed black curve) onto which
a gaussian signal of strength $S$ set equal to the found upper limit (CL=90\%) of 
$BR_{\eta \rightarrow e^+e^-} < 2.5 \times 10^{-6}$ is superimposed (solid red and pink curves).
}  
\label{zoom_eta}
\end{figure}

The direct decay of the $\eta$ meson into a lepton pair ($e^+e^-$ or $\mu^+\mu^-$) can only proceed
through a 2-photon intermediate state.  The \ee\ decay is furthermore strongly suppressed
by helicity conservation.  Calculations based on chiral perturbation theory and quark models
put its branching ratio at $BR_{\eta\rightarrow e^+e^-}^{QCD} \simeq 5\times10^{-9}$ \cite{savage,dorokhov}.
The previous 90~\%~CL upper limit on the $\eta \rightarrow e^+e^-$ decay branch,
obtained from HADES p+p data \cite{hades_p35p}, has been fixed by the 2012 review of the PDG \cite{pdg2012}
at $BR_{\eta\rightarrow e^+e^-} < 5.6\times10^{-6}$.  The present analysis of our p+Nb data allows to set
an improved limit (CL = 90\%) at $2.5\times10^{-6}$ (see Fig.~\ref{zoom_eta}).  Combining the p+p and p+Nb results
with the help of Eq.~(\ref{limit_add}), a final limit of $2.3\times10^{-6}$ can be given, i.e. about
a factor 2.5 lower than the present PDG value, but still a far way above
theoretical predictions \cite{savage,dorokhov}.

\section{Summary and outlook}

Searching for a narrow resonance in dielectron spectra measured with HADES in the reactions 
p (at 3.5~\gev) + p, Nb, as well as Ar (at 1.756~\gevu) + KCl we have established an upper limit at 90$\%$~CL
on the mixing $\epsilon^2=\alpha'/\alpha$ of a hypothetical dark photon $U$ in the mass range
$M_U = 0.02 - 0.6$~\gevcc.  Our UL sets a tighter constraint than the recent WASA search at low masses
excluding to a large extent the parameter space preferred by the muon $g-2$ anomaly.  At higher masses,
already surveyed by the recent KLOE-2 search, our analysis provides complementary information.
We have thus covered for the first time in one and the same experiment a rather broad mass range.
In addition, we have reduced the UL on the direct decay $\eta \rightarrow e^+e^-$ by a factor 2.5 with respect
to the known limit to $2.3 \times 10^{-6}$.  In future experiments at the FAIR facility we expect to be able
to increase our sensitivity by up to one order of magnitude.

\section{Acknowledgments}

The HADES Collaboration gratefully acknowledges the support
by BMBF grants 06DR9059D, 05P12CRGHE, 06FY171, 06MT238 T5,
and 06MT9156 TP5, by HGF VH-NG-330, by DFG EClust
153, by GSI TMKRUE, by the Hessian LOEWE initiative
through HIC for FAIR (Germany), by EMMI GSI,
TU Darmstadt (Germany): VH-NG-823, Helmholtz Alliance HA216/EMMI,
by grant GA CR 13-067595 (Czech Rep.),
by grant NN202198639 (Poland),
Grant UCY/3411-23100 (Cyprus),by CNRS/IN2P3 (France),
by INFN (Italy), and by EU contracts RII3-CT-2005-515876
and HP2 227431.




\begin{thebibliography}{00}

\bibitem{atic} J. Chang \etal, Nature 456 (2008) 362.
\bibitem{pamela} O. Adriani \etal, Nature 458 (2009) 607.
\bibitem{hess1} F. Aharonian \etal, Phys. Rev. Lett. 101 (2008) 261104.
\bibitem{hess2} F. Aharonian \etal, Astron. Astrophys. 508 (2009) 561.
\bibitem{fermi} A. Abdo \etal, Phys. Rev. Lett. 102 (2009) 181101.
\bibitem{ams02} M. Aguilar \etal, Phys. Rev. Lett. 110 (2013) 141102.
\bibitem{wimpdecays1} I.~Cholis, L.~Goodenough, N.~Weiner, Phys. Rev. D 79 (2009) 123505.
\bibitem{wimpdecays2} I.~Cholis \etal, Phys. Rev. D 80 (2009) 123518.
\bibitem{darkmatter} G.~Bertone, D. Hooper, J. Silk, Phys. Rept. 405 (2005) 279.
\bibitem{pdg2012} J.~Behringer \etal\ (Particle Data Group), Phys. Rev. D 86 (2012) 010001.
\bibitem{fayet80} P.~Fayet, Phys. Lett. B 95 (1980) 285.
\bibitem{boehm} C.~Boehm, P. Fayet, Nucl. Phys. B 683 (2004) 219.
\bibitem{fayet04} P.~Fayet, Phys. Rev. D 70 (2004) 023514.
\bibitem{pospelov08} M.~Pospelov, A. Ritz, M. B. Voloshin, Phys. Lett. B 662 (2008) 53.
\bibitem{okun} L.~Okun, Sov. Phys. JETP 56 (1982) 502.
\bibitem{holdom} B.~Holdom, Phys. Lett. B 166 (1986) 196.
\bibitem{arkani} N.~Arkani-Hamed, D.P. Finkbeiner, T.R. Slatyer, N. Weiner, Phys. Rev. D 79 (2009) 015014.
\bibitem{bjorken} J.~Bjorken \etal, Phys. Rev. D 80 (2009) 075018.
\bibitem{batell1} B.~Batell, M. Pospelov, A. Ritz, Phys. Rev. D 79 (2009) 115008.
\bibitem{reece} M.~Reece, L.-T. Wang, JHEP 0907 (2009) 051.
\bibitem{landsberg} L. Landsberg, Phys. Rept. 128 (1985) 301.
\bibitem{mamiA1} H.~Merkel \etal\ (A1 Collaboration), Phys. Rev. Lett. 106 (2011) 251802.
\bibitem{apex} S.~Abrahamyan \etal\ (APEX Collaboration), Phys. Rev. Lett. 107 (2011) 191804.
\bibitem{sindrum} R.~Meijer Drees \etal (SINDRUM I Collaboration), Phys. Rev. D 45 (1992) 1439.
\bibitem{wasa} P.~Adlarson \etal\ (WASA-at-COSY Collaboration), Phys. Lett. B 726 (2013) 187.
\bibitem{kloe-2a} F.~Archilli \etal\ (KLOE-2 Collaboration), Phys. Lett. B 706 (2012) 251.
\bibitem{kloe-2b} D.~Babuski \etal\ (KLOE-2 Collaboration), Phys. Lett. B 720 (2013) 111.
\bibitem{gninenko} S.~Gninenko, Phys. Rev. D 85 (2012) 055027.
\bibitem{muong2} G.W.~Bennett \etal\ (Muon (g-2) Collaboration), Phys. Rev. D 73 (2006) 072003.
\bibitem{pospelov09} M.~Pospelov, Phys. Rev. D 80 (2009) 095002.
\bibitem{endo} M.~Endo, K.~Hamaguchi, G. Mishima, Phys. Rev. D 86 (2012) 095029.
\bibitem{hades_p35p} G.~Agakishiev \etal\ (HADES~Collaboration), Eur. Phys. J. A 48 (2012) 64.
\bibitem{hades_pNb} G.~Agakishiev \etal\ (HADES~Collaboration), Phys. Lett. B 715 (2012) 304.
\bibitem{hades_pi0_eta} G.~Agakishiev \etal\ (HADES~Collaboration), Phys. Rev. C 88 (2013) 024904.
\bibitem{hades_arkcl} G.~Agakishiev \etal\ (HADES~Collaboration), Phys. Rev. C 84 (2011) 014902.
\bibitem{batell2} B.~Batell, M. Pospelov, A. Ritz, Phys. Rev. D 80 (2009) 095024.
\bibitem{pluto} F.~Dohrmann \etal, Eur. Phys. J. A 45 (2010) 401.
\bibitem{berghauser} H.~Bergh\"{a}user \etal, Phys. Lett. B 701 (2011) 562.
\bibitem{NA60} R.~Arnaldi \etal\ (NA60~Collaboration) Phys. Lett. B677 (2009) 260.
\bibitem{pena} G.~Ramalho, M.T. Pena, A. Stadler, Phys. Rev. D86 (2012) 093022.
\bibitem{hades_tech} G.~Agakishiev \etal\ (HADES~Collaboration), Eur. Phys. J. A 41 (2009) 243.
\bibitem{GiBUU} J. Weil, H. van Hees, U. Mosel, Eur. Phys. J. A 48 (2012) 111.
\bibitem{cowan} G.~Cowan, K. Cranmer, E. Gross, O. Vitells, Eur. Phys. J. C 71 (2011) 1554.
\bibitem{rolke} W.A.~Rolke, A.M. Lopez, J. Conrad, Nucl. Inst. Meth. Phys. Res. A 551 (2005) 493.
\bibitem{root} R.~Brun, F. Rademaker, Nucl. Instr. and Meth. A 389 (1997) 81. See also http://root.cern.ch/. 
\bibitem{helene} O.~Helene, Nucl. Instr. and Meth. Phys. Res. A 390, (1997) 383.
\bibitem{BaBar} B.~Aubert \etal\ (BaBar~Collaboration) (2009), arXiv:0902.2176.
\bibitem{savage} M.~Savage, M.~Luke, M.B.~Wise, Phys. Lett. B 291 (1992) 481.
\bibitem{dorokhov} A.E.~Dorokhov, M.A.~Ivanov, Phys. Rev. D 75 (2007) 114007.
\end{thebibliography}
\end{document}